\documentclass[natbib=false,acmsmall,nonacm,screen]{acmart}
\usepackage{babel}
\usepackage[utf8]{inputenc}
\usepackage{wrapfig}
\usepackage{csquotes}
\usepackage{subcaption}
\usepackage{cleveref}
\usepackage[shortlabels]{enumitem}
\usepackage{listings}
\usepackage{multirow}
\usepackage{xcolor,colortbl}
\usepackage{array}

\newcommand{\STAB}[1]{\begin{tabular}{@{}c@{}}#1\end{tabular}}

\definecolor{amaranth}{rgb}{0.9, 0.17, 0.31}
\definecolor{tangerine}{rgb}{0.95, 0.52, 0.0}

\definecolor{codegreen}{rgb}{0,0.6,0}
\definecolor{codegray}{rgb}{0.5,0.5,0.5}
\definecolor{codepurple}{rgb}{0.58,0,0.82}

\lstdefinestyle{mystyle}{
    commentstyle=\color{codegreen},
    keywordstyle=\color{magenta},
    numberstyle=\tiny\color{codegray},
    stringstyle=\color{codepurple},
    basicstyle=\ttfamily\footnotesize,
    breakatwhitespace=false,         
    breaklines=true,                 
    captionpos=b,                    
    keepspaces=true,                 
    numbers=left,                    
    numbersep=5pt,                  
    showspaces=false,                
    showstringspaces=false,
    showtabs=false,                  
    tabsize=2
}

\authorsaddresses{}

\acmBooktitle{}
\begin{document}

\newcommand{\FIGUREWIDTH}{0.326\textwidth}

\title{Towards Effective and Efficient Padding Machines for Tor}
\subtitle{The good, the bad, and the ugly. \today}

\author{Tobias Pulls}
\email{tobias.pulls@kau.se}
\affiliation{%
  \institution{Computer Science, Karlstad University, Sweden}
  \city{Karlstad}
}

\renewcommand{\shortauthors}{Tobias Pulls}

\begin{abstract}
	Tor recently integrated a circuit padding framework for creating padding
	machines: defenses that work by defining state machines that inject dummy
	traffic to protect against traffic analysis attacks like Website
	Fingerprinting (WF) attacks. In this paper, we explore the design of
	effective and efficient padding machines to defend against WF attacks.
	Through the use of carefully crafted datasets, a circuit padding simulator,
	genetic programming, and manual tuning of padding machines we explore
	different aspects of what makes padding machines effective and efficient
	defenses. Our final machine, named Interspace, is probabilistically-defined
	with a tweakable trade-off between efficiency and effectiveness against the
	state-of-the-art deep-learning WF attack Deep Fingerprinting by Sirinam et
	al.~\cite{df}. We show that Interspace can be both more effective and
	efficient than WTF-PAD by Juarez et al.~\cite{JuarezIPDW16}, the padding
	machine that inspired the design of Tor's circuit padding framework. We end
	this paper by showing how Interspace can be made less effective, identifying
	the promising tactic of probabilistically defined padding machines, and
	highlighting the need to further explore this tactic in more complex
	defenses.
\end{abstract}

\maketitle

\section{Introduction}
Tor~\cite{tor} recently integrated a circuit padding framework for circuit-level
defenses in the form of padding
machines\footnote{\url{https://gitweb.torproject.org/torspec.git/tree/padding-spec.txt}}.
The framework is thoroughly
documented~\footnote{\url{https://gitweb.torproject.org/tor.git/tree/doc/HACKING/CircuitPaddingDevelopment.md}}
and a simulator for the framework has been
developed\footnote{\url{https://github.com/pylls/circpad-sim}}. Moving forward,
we expect that researchers will use the circuit padding framework to build new
padding machines with the goal of defending against a range of traffic analysis
attacks. One such attack---that was one of the motivations behind the design of
the framework---is a Website Fingerprinting (WF)
attack~\cite{cheng1998traffic,DBLP:conf/sp/SunSWRPQ02,Hintz02,DBLP:conf/ccs/LiberatoreL06,HerrmannWF09,PanchenkoNZE11}.
In a WF attack, a passive local eavesdropper (e.g., a guard relay or an ISP)
observes the encrypted traffic between a Tor Browser client and the Tor network
and attempts to infer the destination based on patterns in the encrypted network
traffic.

This paper is dedicated to the evaluation of WF attacks and defenses involving
padding machines in the circuit padding framework and accompanying simulator. We
present a dataset and evaluation method for the circuit padding simulator
structured for \emph{webpage-to-website} fingerprinting, building upon the work
of Cai et al.~\cite{touching} and Panchenko et al.~\cite{cumul}
(Section~\ref{sec:dataset}). Using the state-of-the-art Deep Fingerprinting (DF)
attack by Sirinam et al.~\cite{df}, we show that DF is highly successful in
webpage-to-website fingerprinting, that the safest security level for TB is
quite different from the others, and that the start of traces is the most useful
for DF (Section~\ref{sec:resultsdf}).

Using genetic programming together with the circuit padding simulator, we
evolved padding machines from February until June 2020, where the best machine
achieve 0.57 max recall\footnote{We focus on max recall because of the results
of Pulls and Dahlberg~\cite{wfwo}, assuming that an attacker can use so-called
website oracles to reduce the false positive rate of its WF classifier.} with
0.52 precision and 206\% overhead (Section~\ref{sec:evolved}). Next, we refine
the evolved machines by manual pruning and tuning, resulting in two machines:
Spring and Interspace. Spring is a simplified machine from the best evolved
machine. Interspace in turn is based on Spring with the main change of being a
probabilistically defined machine, similar to how ScrambleSuit (and obfs4)
works~\cite{DBLP:conf/wpes/WinterPF13}. The Interspace machine has a max recall
slightly above 0.3 with 230\% overhead. In the same setting, the WTF-PAD defense
by Juarez et al.~\cite{JuarezIPDW16}---that inspired the design of Tor's circuit
padding framework---has a max recall of about 0.7 at 178\% overhead. Interspace
is tweakable, allowing an effectiveness-efficiency trade-off. At the same
overhead as WTF-PAD, Interspace has a max recall of 0.5, and can provide 0.7 max
recall at 150\% overhead (Section~\ref{sec:tuned}).

Section~\ref{sec:discussion} discusses the good, the bad, and the ugly of our
results. The good was earlier mentioned: Interspace is an effective machine in
our setting, tweakable, and built for Tor's circuit padding framework. In terms
of the bad, we are basically brute-forcing the design of the machines against a
particular deep learning attack through simulation. Also, we completely ignore
time, since DF does not use time and time in the simulator is unrealistically
accurate. For the ugly, we show that he effectiveness of Interspace could be
reduced by significantly increasing the training data available to DF through
repeated simulation of its defense on our collected traces (in a sense, changing
our experimental setting), reaching a max recall above 0.6.

We conclude this paper in Section~\ref{sec:conclusions}. Interspace, being a
probabilistically defined machine, could be made less effective with more
training data. However, we show (again) that DF struggles once deprived of being
able to train on the exact defense used by a target. This tactic should be
further explored for more complex defenses.

\section{The Goodenough Dataset}
\label{sec:dataset}
We set out to create a dataset that better reflects the challenges of an
attacker than the typical datasets used in the evaluation of WF attacks. The
dataset consists of 10,000 monitored samples and 10,000 unmonitored samples
(balanced, so not taking into account client or network base rate). The
monitored samples represent 50 classes of popular websites taken from the Alexa
toplist (all within Alexa top-300 at the time of collection). For each
website/class, we selected ten webpages of that website to represent that class,
hence \emph{webpage-to-website} fingerprinting. For example, for the website
\url{reddit.com}, we selected ten URLs to popular subreddits such as
\href{https://www.reddit.com/r/wholesomememes/}{r/wholesomememes} and
\href{https://www.reddit.com/r/Wellthatsucks/}{r/Wellthatsucks/}. Similarly, for
\url{wikipedia.org}, we selected articles such as
\url{https://en.wikipedia.org/wiki/Dinosaur} and
\url{https://en.wikipedia.org/wiki/Fossil}. The full list of websites and
webpages are available as part of the dataset. We collected 20 samples per
webpage, resulting in $50x10x20=10,000$ monitored samples.

Now to the key idea of the dataset: to evaluate WF attacks based on classifiers
that need training, we split the dataset into training, validation, and testing
with a 8:1:1 ratio. This split is done \emph{per webpage}, such that the
classifier never gets to train on the same webpage as is used for validation or
testing. This trains the classifier to identify the \emph{common structure} that
makes up different webpages belonging to the same website/class. 

\begin{figure*}[ht]
	\centering
	\includegraphics[width=0.75\textwidth]{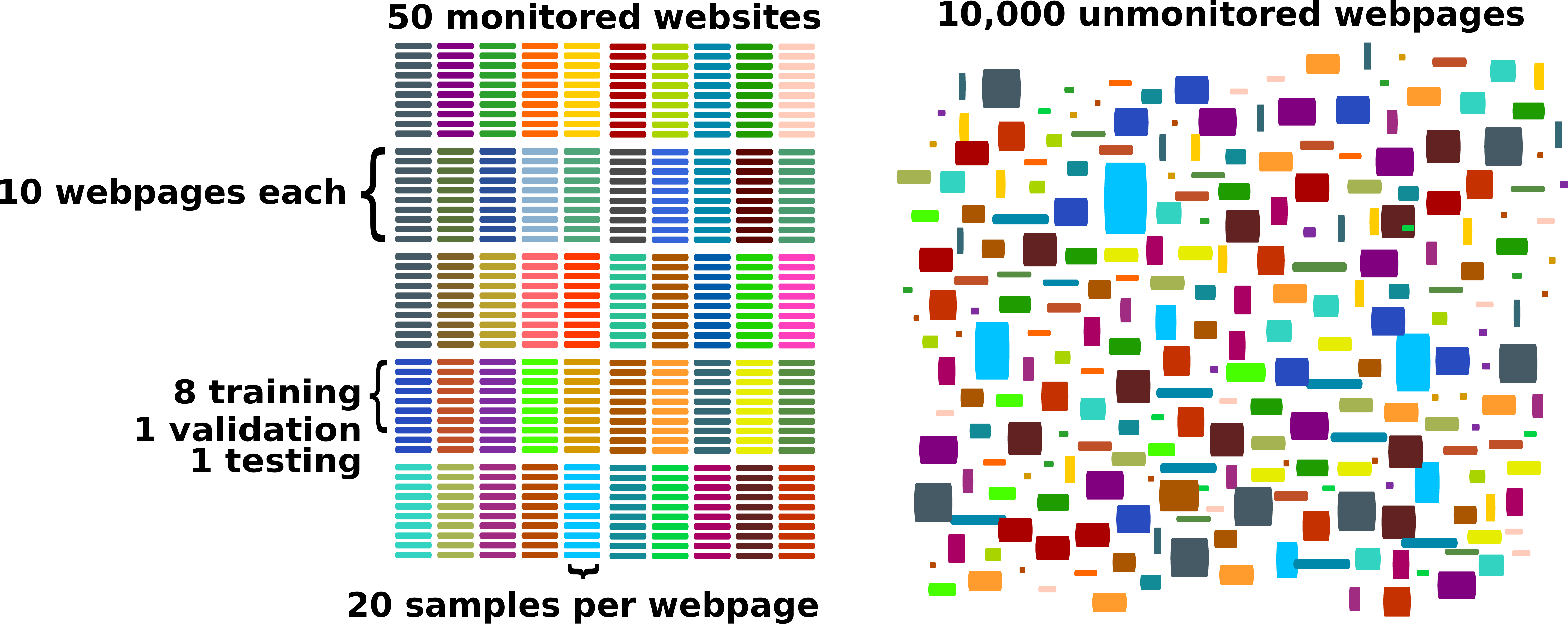}
	\caption{A visualization of the structure of the Goodenough dataset.}
	\label{fig:dataset}
\end{figure*}

Figure~\ref{fig:dataset} visualizes the structure of our dataset, including the
unmonitored samples, which consists of 10,000 unmonitored webpages collected
from \url{reddit.com/r/all} (top last year). We made sure to exclude webpages of
monitored websites, which include self-hosted images at Reddit. We also excluded
direct image links, since they are too distinct to the monitored webpages, and
links to YouTube and Twitter that have a tendency of sporadically blocking
access from the Tor network.

\subsection{Security Levels}
The dataset exists in three versions with one for each security level in Tor
Browser: standard, safer, and safest. The security level determines if a number
of features in the browser are enabled or not, where one of if not the most
notable feature is JavaScript. JavaScript is completely disabled at the safest
security level. Three versions of the dataset allows us to better take into
account another important aspect often overlooked in the evaluation of WF
attacks and defenses.

\subsection{Key Collection Details}
\label{sec:collection-details}
All samples in the final dataset were gathered with Tor Browser 9.0.2 in
February, 2020, from Karlstad University, Sweden (connected to
SUNET\footnote{\url{https://www.sunet.se}}) using freely available
tools\footnote{\url{https://github.com/pylls/padding-machines-for-tor/tree/master/collect-traces}}
with ample compute and network resources. We also collected the full dataset in
December, 2019, and January, 2020. No differences in results between the three
different points in time were noticed, beyond having to update a few webpages
that disappeared (primarily news articles). We therefore opt to only use the
February dataset for the rest of this paper unless stated otherwise.

Each sample consists of one log and two traces. The log is the modified tor log
from Tor Browser using the
circpad-sim\footnote{\url{https://github.com/mikeperry-tor/tor/commits/circpad-sim-v4}}
modifications, generated by having 140 concurrent docker containers of headless
Tor Browser visiting webpages for one minute. In Tor Browser, we disabled guards
(\texttt{UseEntryGuards 0}) and altered the security level. Each visit used a
dedicated copy of Tor Browser with a fresh copy of the consensus (to reduce the
load on the network). Therefore, each sample was collected on its own circuit.

From each log, we generated the client trace from circpad-sim using
\texttt{torlog2circpadtrace.py}\footnote{\url{https://github.com/pylls/circpad-sim/blob/master/torlog2circpadtrace.py}}
with default parameters (note the 10,000 maximum length per circuit to extract).
We also include corresponding \emph{simulated} relay traces, generated using
\texttt{simrelaytrace.py}\footnote{\url{https://github.com/pylls/circpad-sim/blob/master/simrelaytrace.py}}
with default parameters. Please note that accuracy of the timestamps in the
traces are \emph{unrealistically accurate} for \emph{any} possible network
attacker. This is because the timestamps are generated by tor with nanosecond
precision \emph{before} the cells are ultimately packaged into TCP/IP packets
(inside a TLS record) and transmitted.

We made efforts to try to ensure that the monitored logs and traces actually
correspond to the intended webpages, discarding any traces where tor/Tor Browser
may have used two or more circuits to load content for the visit (for unknown
reasons). Less care was taken for unmonitored samples to better represent
realistic background traffic (that may use multiple circuits for unknown yet
sporadically observed reasons).

\subsection{Related Work}
Both the work of Panchenko et al.~\cite{cumul} and Cai et al.~\cite{touching}
inspired how our dataset is structured and its implied method of use. Both works
use traditional machine learning techniques with manually engineered features,
significantly reducing the effectiveness of attacks in their evaluations.

Cai et al. differentiate between websites and webpages, where an attacker's goal
is to identify a visit to any webpage of a website. They note that webpages of
websites commonly share structure (``templates''), and that an attacker can
attempt to detect that structure regardless of exact webpage visited by a
victim. This is identical to our work. However, their dataset is small,
consisting of two monitored websites, with an unclear number of unmonitored
websites (open world is stated though). They also focus more on modelling
complete browsing sessions by victims, at a time when many of the advances in WF
attacks were still to see the light of day.

Panchenko et al. also differentiate between websites and webpages, but in
general, train on the same webpage as they later classify. They did note one
attempt of only training on a subset of the webpages later used for testing, in
the closed world, where the performance of their classifier degraded
significantly. This is our default setting, but in the open world. We used
reddit to gather our unmonitored dataset, while Panchenko et al. in part used
Twitter.

In gist, how our dataset is structured and its method of evaluation is based
upon the work of Panchenko et al.~\cite{cumul} and Cai et al.~\cite{touching},
but our results and extended experiment design benefit from the recent advances
in automatic feature engineering using deep learning. 

\section{Results using Deep Fingerprinting}
\label{sec:resultsdf}
For all the following results, we extracted the first 5,000 cells\footnote{Note
that cells make up a subset of the 10,000 maximum events extracted from logs, as
noted in Section~\ref{sec:collection-details}.} and their directions from the
dataset for each sample, \emph{completely discarding time}, because it is
unrealistically precise. We use the state-of-the-art Deep Fingerprinting (DF)
attack by Sirinam et al.~\cite{df}, ported to the PyTorch library. Deep-learning
based attacks (like DF) free us from the burden of manual feature engineering
and have been shown to be superior to traditional machine-learning WF attacks in
many
settings~\cite{DBLP:journals/popets/BhatLKD19,DBLP:journals/popets/OhSH19,df}.

Because the dataset is of modest size---20,000 traces per security level---it is
feasible to perform ten-fold cross-validation when using DF and a modern GPU. For
each fold, we change the webpages used to train, validate, and test the
monitored websites as well as the unmonitored webpages deterministically, never
training on any webpage used for validation or testing. 

\subsection{Webpage-to-Website Fingerprinting}
\begin{figure}[ht]
  \centering
  \includegraphics[width=0.35\textwidth]{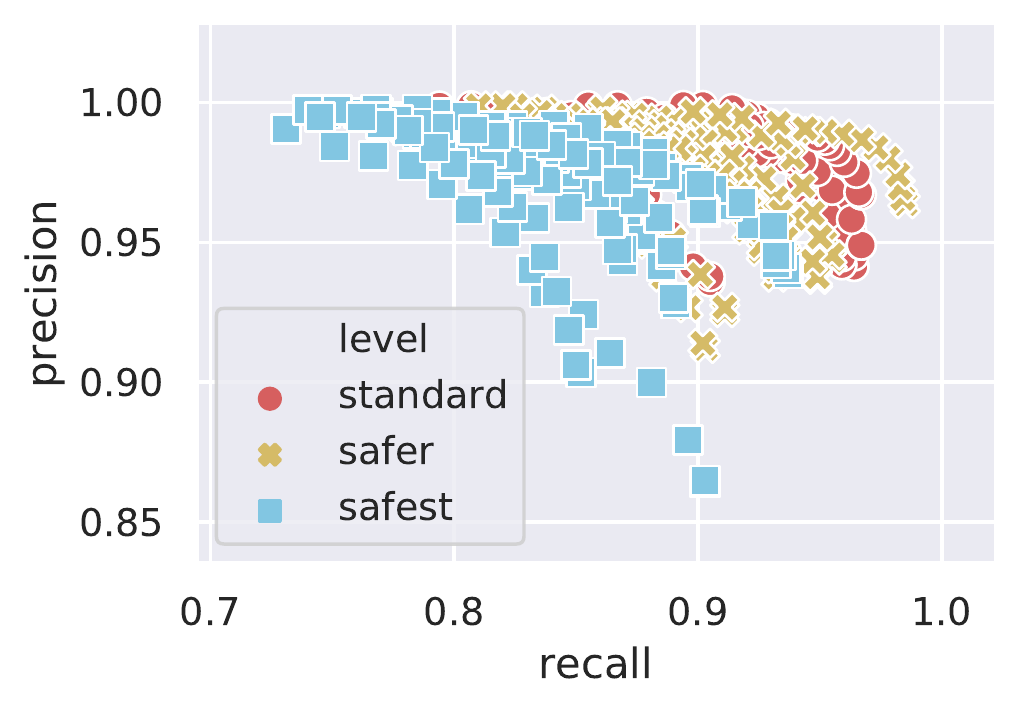}
  \caption{10-fold cross-validation with 16 thresholds for the three different security levels in Tor Browser.}
  \label{fig:10-fold}
\end{figure}

Figure~\ref{fig:10-fold} shows the results of performing webpage-to-website
fingerprinting for all three security levels of Tor Browser. Each dot represents
a possible precision-recall trade-off---based on the probability threshold for
the monitored classes from the final softmax in DF---for ten-fold
cross-validation. The DF attack is highly successful. We see slightly better
protection offered by the safest security level on average (note the axes), with
no difference between standard and safer.

\subsection{Classification Across Security Levels}
\begin{figure*}[h]
	\centering
		\begin{subfigure}[b]{\FIGUREWIDTH}
		\includegraphics[width=1\textwidth]{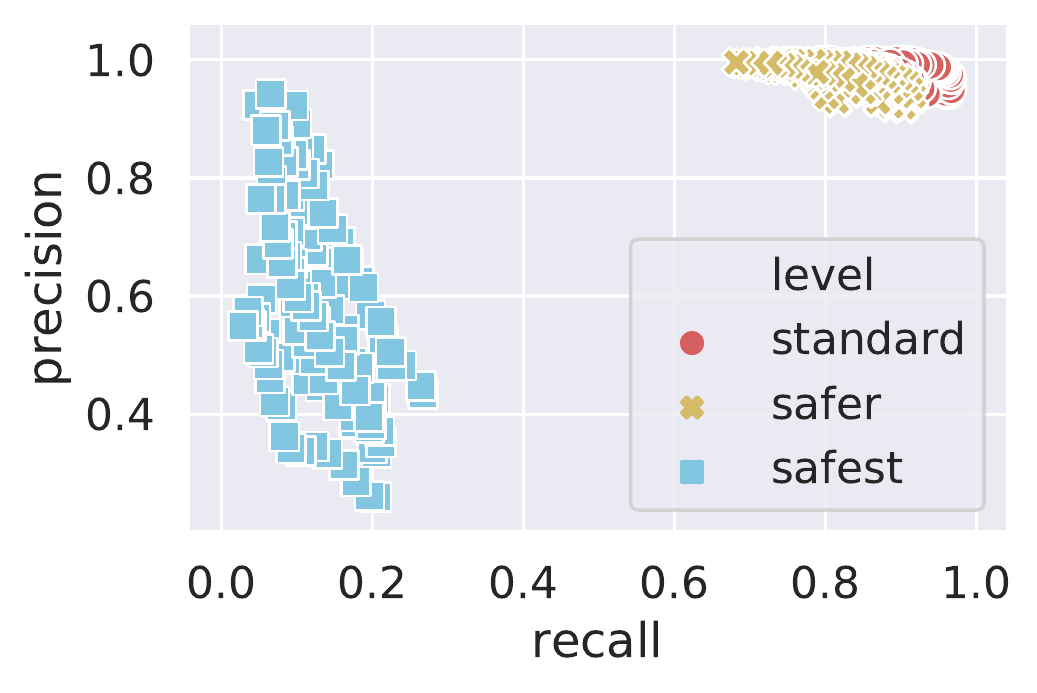}
		\caption{Trained on standard}
		\label{fig:levels:standard}
		\end{subfigure}
		\begin{subfigure}[b]{\FIGUREWIDTH}
		\includegraphics[width=1\textwidth]{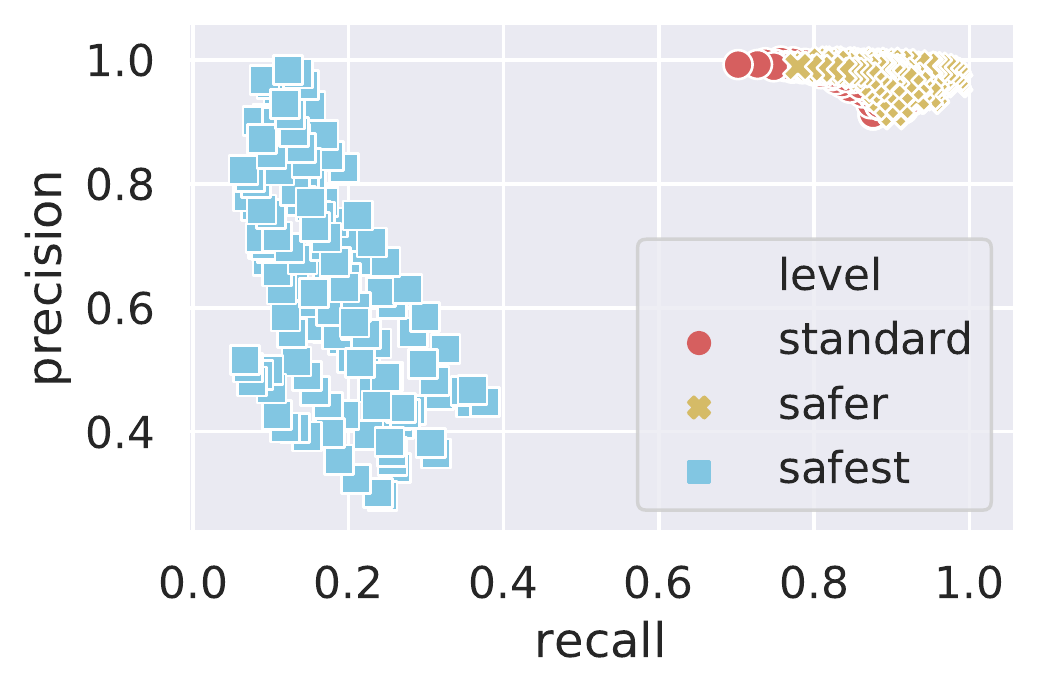}
		\caption{Trained on safer}
		\label{fig:levels:safer}
		\end{subfigure}
		\begin{subfigure}[b]{\FIGUREWIDTH}
		\includegraphics[width=1\textwidth]{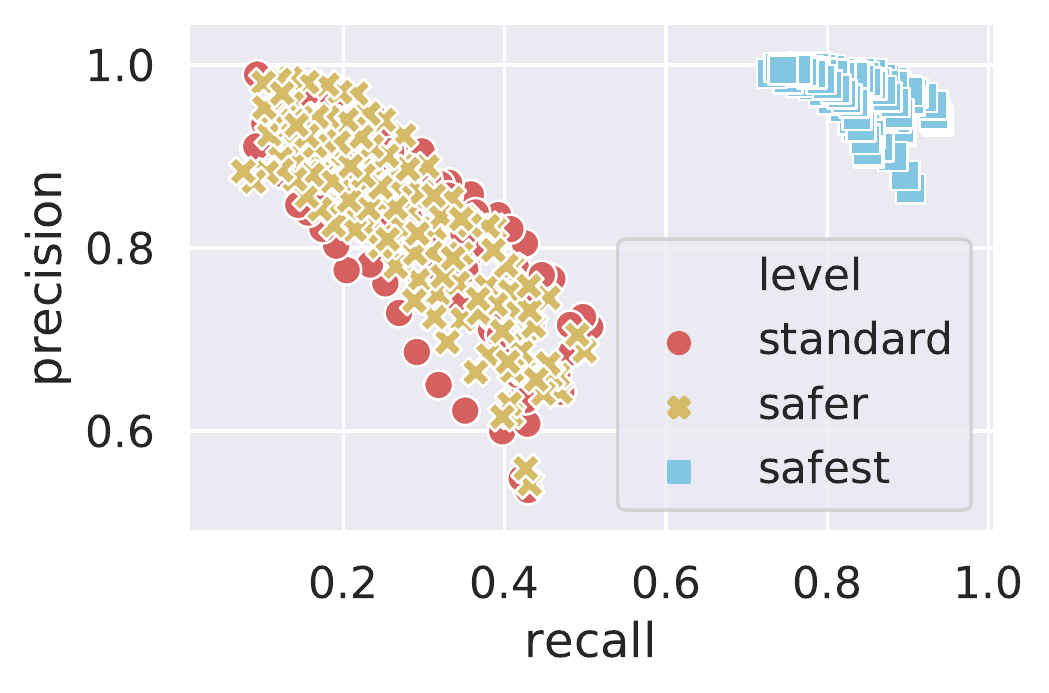}
		\caption{Trained on safest}
		\label{fig:levels:safest}
		\end{subfigure}
		\caption{Precision-recall metrics on datasets collected using the three security levels in Tor Browser, when the model is trained on data belonging to a particular security level.}
		\label{fig:levels}
\end{figure*}

Figure~\ref{fig:levels} shows the impact of cross-level classification, where a
model trained on a dataset generated by one security level of Tor Browser is
used to classify all three levels. We see a significant difference when it comes
to the safest security level: the impact on webpage-to-website fingerprinting is
huge. There is no significant difference between standard and safer, similar to
the results in Figure~\ref{fig:10-fold}.

\subsection{Estimating Defense Priorities}
\label{sec:dataset:zero}
We can estimate the relative importance of different parts of a trace by zeroing
cells, removing a variable amount of information from the attacker. In the
following analysis we only consider the maximum recall by setting the threshold
in DF to 0 for each of the ten-folds per measurement.

\begin{figure*}[ht]
	\centering
		\begin{subfigure}[b]{\FIGUREWIDTH}
			\includegraphics[width=1\textwidth]{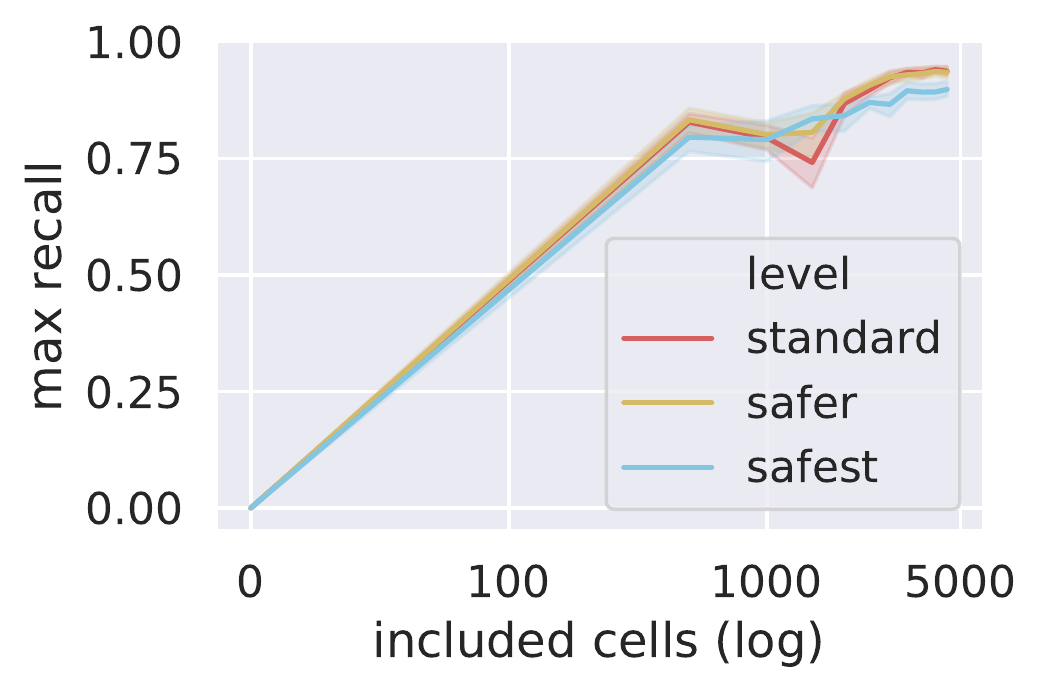}
			\caption{Increasingly including}
			\label{fig:zero:inc}
		\end{subfigure}
		\begin{subfigure}[b]{\FIGUREWIDTH}
			\includegraphics[width=1\textwidth]{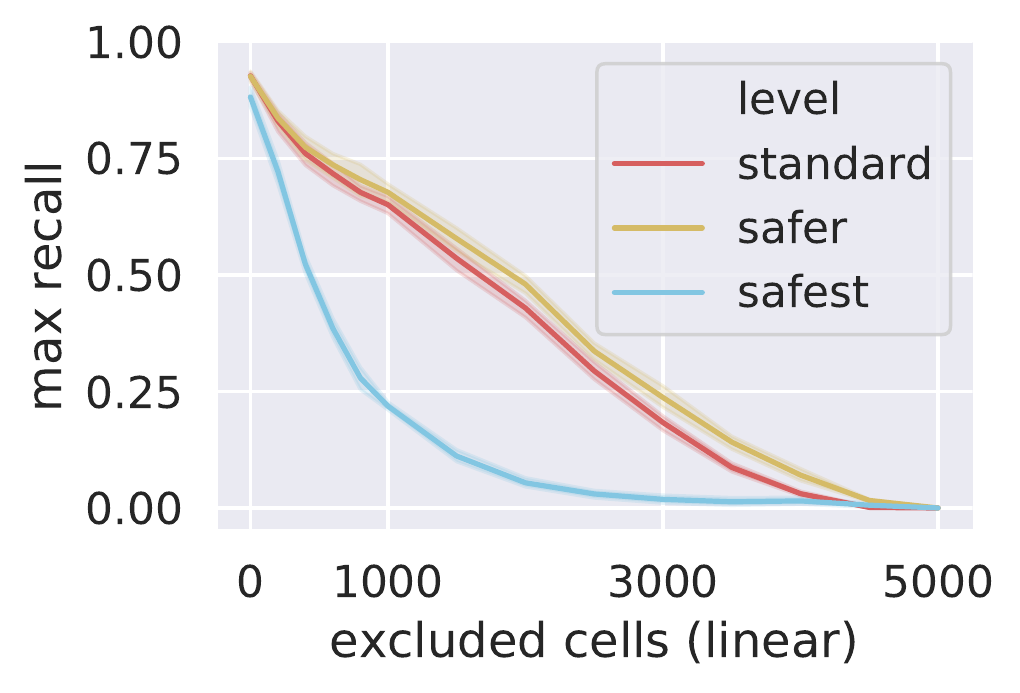}
			\caption{Increasingly excluding}
			\label{fig:zero:dec}
		\end{subfigure}
		\caption{The maximum recall when including or excluding increasingly
		more cells of traces. Based on the dataset collected in January (took
		about a week to produce with ten folds, cannot justify costs when no
		other scenario offers any noticeable difference).}
		\label{fig:zero}
\end{figure*}

Starting from an empty trace, Figure~\ref{fig:zero:inc} includes increasingly
more cells in traces. Note the logarithmic x-axis. The figure shows a steep
increase in recall with a few hundreds of cells, needing only about 100 cells to
hit 0.5 recall. No significant difference between security levels, with the
exception of a small dip for standard and safer around 1,500 cells.

Figure~\ref{fig:zero:dec} starts with a full trace and excludes increasingly
more cells. Note the linear x-axis. Comparing the results to
Figure~\ref{fig:zero:inc}, it is clear that the start of a trace has the most
utility for an attacker: the first 100 cells are about as useful as the last
3,000 cells for standard and safer security levels. This is in line with the
results of Mathews et al.~\cite{featurediscovery}. However, the most striking
result is the impact on the safest security level. By excluding the first 1,000
cells the maximum recall is cut to about 0.25, approximately a third of the
recall for the other security levels.

The degree to which the safest security level stands out in
Figure~\ref{fig:zero:dec} can be explained in part with a closer look at the
dataset. The safest dataset consists of only 1,534 full traces (5,000 cells),
with an average 1,478 and mean 950 of cells per trace.

\section{Evolved Machines}
\label{sec:evolved}
During the first half of 2020 we used genetic programming together with the
circuit padding simulator to evolve machines. We evolved machines for several
months of wall-time using an AMD Ryzen 7 2700X and NVIDIA GeForce RTX 2070.

\subsection{Method}
\label{sec:evolved:method}

We experimented early with a number of different fitness functions (including
information-theoretic ones based on trace collisions), but landed on
$1-\texttt{recall}$, where recall is computed using DF with threshold $0.0$.
That is, we aim to minimize the maximum recall of DF against the evolved
machines. Recall is used because it captures when the client visits a monitored
page, and it tells us how many out of all visits by the client the classifier
accurately detects. More importantly, as a metric, it ignores how the classifier
performs against unmonitored testing traces. DF does not use time, so it is
appropriate to use with the circuit padding simulator. It might seem overkill to
use deep learning, but our execution time was dominated by simulating the
evolved machines. In our implementation, to reach parity between simulating
traces and computing the fitness function (deep learning) using a modest NVIDIA
GeForce RTX 2070 we would need in the order of a 100 CPU cores.

We used a small population size of ten machines, where the initial population
was randomly generated. This might seem small, but each machine has to be
simulated against 20,000 traces, taking significant time. We experimented
relatively briefly with different population sizes, but found no improvement in
larger sizes. We used the full dataset of 20,000 traces at the safest security
level (the smallest and easiest to protect, see Section~\ref{sec:dataset:zero})
to make maximum recall as good as possible of a fitness function.

Given a population of machines and their fitness determined based on their max
recall with DF, the next generation was evolved as follows. First, we selected
an \emph{elitist fraction} of the most fit machines and copied them over as-is.
For most of our runs, this was only the single best machine. We briefly
experimented with keeping a \emph{diversity fraction} of randomly selected
machines, but opted to disable this due to lack of initial results. For the
remaining slots in the population (typically nine), we evolved new machines
using crossover and mutation.  For crossover, we performed single-point
crossover between \emph{complete states} of machines. For mutation, with some
probability, we mutated each part of each state (the inter-arrival distribution,
the length distribution, and the state transitions). Parents were selected at
random from the previous generation, weighted by fitness. The client and relay
parts of machines never interacted; each evolved independently.

Each machine actually consists of a pair of machines: one machine running on the
client and one on the relay (we target the second hop, i.e., the middle relay).
In particular, we do not evolve \emph{symmetric} machines (as in, e.g.,
WTF-PAD), so the client and relay logic may be completely different. A single
machine consists of four states (compared to three in WTF-PAD), but only uses
distributions for length- and time-sampling (so no histograms) to limit the
search space. It is also known from WTF-PAD that tailoring histograms is a
significant challenge for deployment (or even evaluation across datasets). The
circuit padding framework enables us to put both absolute and relative overhead
limits for padding \emph{circuit} and machine. We used generous limits, with the
final machines allowing 1,000 cells in absolute terms, that are then followed by
a 50\% overhead as a percentage of total traffic. Note that for evaluating the
fitness of machines we use the first 5,000 cells of each trace. Finally, all
machines start as soon as streams are attached to the circuit. Due to how we
collected our dataset, the first stream directly carries data to connect to the
destination website.

\subsection{Results}
\label{sec:evolved:results}
During each month---from February until June 2020---we looked at the most fit
machines and made minor tweaks and/or fixes to our software for evolving
machines. Table~\ref{table:evolved:summary} gives a summary of the most fit
machines each month and their bandwidth overheads in total and per direction
(from the client's perspective). For reference, on unprotected traces DF gets
0.88 max recall and 0.93 precision. From the table, we note that the June
machine is basically useless, despite having the highest total bandwidth
(242\%). In comparison, from the evolved machines, the best machine is April
with a max recall of 0.57, 0.52 precision, and only 206\% overhead.

\begin{table*}[h]
	\caption{The five best evolved machines from February to June with their
		maximum recall and corresponding precision against the safest dataset.
		Average bandwidth (BW) is shown in total (where 100\% means no overhead)
		as well as per direction (fraction of total in parenthesis).}
	\centering
	\label{table:evolved:summary}
	\begin{tabular}{l c c c c c}
	\textbf{Machine} & \textbf{Max Recall} & \textbf{Precision} & \textbf{Total BW} & \textbf{Sent BW (total)} & \textbf{Recv BW (total)} \\
	\midrule February & 0.79 & 0.79 & 173\% & 817\% (48\%) & 100\% (52\%) \\
	March & 0.86 & 0.79 & 173\% & 816\% (48\%) & 100\% (52\%) \\
	April & 0.57 &  0.52 &206\% & 396\% (19\%) & 185\% (81\%) \\
	May & 0.71 & 0.64 & 182\% & 439\% (24\%) & 153\% (76\%) \\
	June & 0.89 & 0.84 & 242\% & 597\% (25\%) & 202\% (75\%) \\
	\end{tabular}
\end{table*}

\begin{figure*}[h]
	\centering
		\begin{subfigure}[b]{\FIGUREWIDTH}
			\includegraphics[width=1\textwidth]{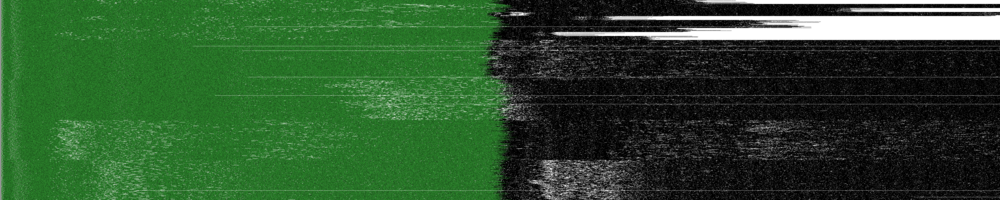}
			\caption{February}
			\label{fig:evolved:february}
		\end{subfigure}
		\begin{subfigure}[b]{\FIGUREWIDTH}
			\includegraphics[width=1\textwidth]{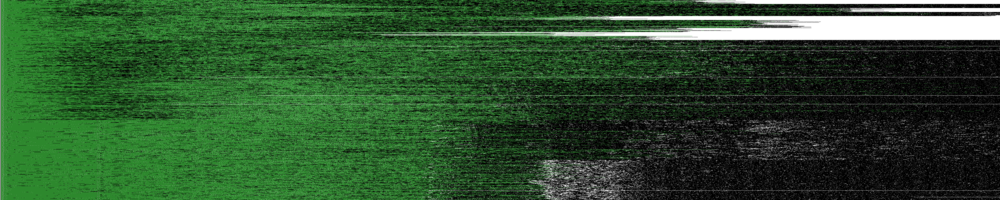}
			\caption{March}
			\label{fig:evolved:march}
		\end{subfigure}
		\begin{subfigure}[b]{\FIGUREWIDTH}
			\includegraphics[width=1\textwidth]{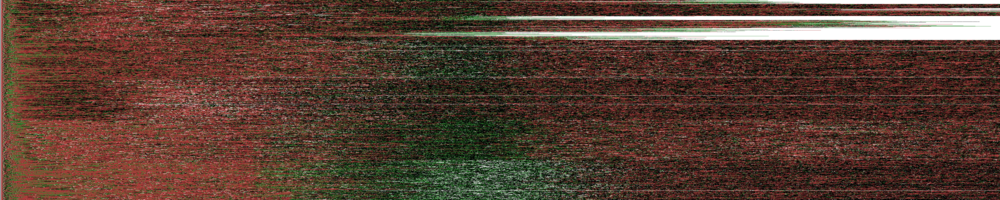}
			\caption{April}
			\label{fig:evolved:april}
		\end{subfigure}
		\begin{subfigure}[b]{\FIGUREWIDTH}
			\includegraphics[width=1\textwidth]{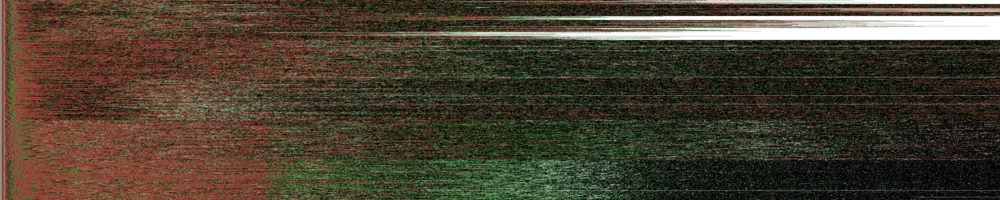}
			\caption{May}
			\label{fig:evolved:may}
		\end{subfigure}
		\begin{subfigure}[b]{\FIGUREWIDTH}
			\includegraphics[width=1\textwidth]{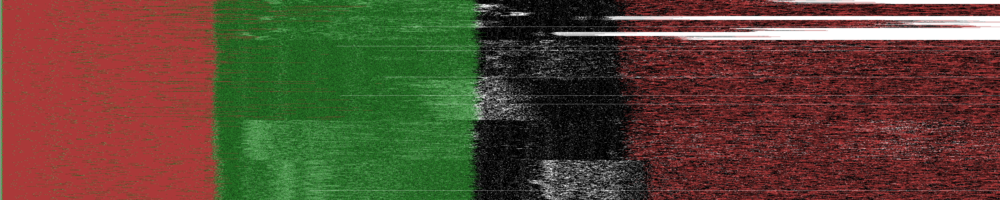}
			\caption{June}
			\label{fig:evolved:june}
		\end{subfigure}
		\caption{Color visualizations of a selection of the traces for selected
		evolved machines. Black represents received nonpadding and white sent
		nonpadding. Red shows received padding and green sent padding.
		Note the type of padding and the importance of padding limits.}
		\label{fig:evolved}
\end{figure*}

From the results in Table~\ref{table:evolved:summary} and details above one
might wonder how come we see such varying results between months. The answer is
simple: bugs in our evaluation code together with the importance of overhead
limits and time spent evolving. To help explain how the machines changed over
time, Figure~\ref{fig:evolved} visualizes the first 200 defended traces for each
selected evolved machine, using colors to differentiate between  sent nonpadding
(white), sent padding (green), received nonpadding (black), and received padding
(red) cells. The completely white connected parts to the top-right of each
figure is empty traces (technically transparent but white due to medium).

Starting with February, we see our first results. From
Figure~\ref{fig:evolved:february} the evolved tactic of the machine is clear:
send as much padding as possible up until the overhead limits kick in. At this
early stage, the allowed padding was set to 2,000 cells, which is slightly less
than half the length of the trace (max 5,000 cells). Also, the fitness function
was set to the recall at threshold 0.5, making the fitness calculation less
stable. Notably, the defense completely ignores any padding from the relay to
the client (received padding, red). This was due to a bug in our use of the
circpad simulator (padding to the guard instead of the middle relay, which is
not supported).

The results from March builds upon February by changing the fitness to the
\emph{minimal} recall (threshold 1.0). While this resulted in worse fitness
(higher max recall), we note from Figure~\ref{fig:evolved:march} that the
padding is now better distributed throughout the traces.

For April we see significant improvements in both
Table~\ref{table:evolved:summary} and Figure~\ref{fig:evolved:april}. We fixed
the bug that prevented any padding from the relay to the client (received
padding, red), set the fitness function to consider the \emph{maximum} recall
(threshold to 0.0), and made the padding limits more strict (1,000 padding cells
then at most 50\% padding). Both sent and received padding are distributed
throughout the traces, achieving 0.57 max recall. We will return to this machine
in more detail later.

For the May machine we restarted evolving machines from a clean slate with the
hope of finding other tactics for padding. Table~\ref{table:evolved:summary}
shows weaker protection than April, but still a notable improvement over no
defense, and less overhead. As we can see in Figure~\ref{fig:evolved:may}, the
received overhead is significantly reduced (note the total as well as received
bandwidth reduction), in particular towards the end of the traces.

The last evolved machine is from June, where we restarted from scratch and fixed
yet another bug. The generated c-code for machines did not correctly set
\texttt{dist\_added\_shift\_usec}: a value added to the sampled delay
distribution for deriving the final padding delay. While this does not
invalidate our prior results, it makes our search space significantly bigger due
to being such a central parameter (up to eight states have this parameter and
large, randomly generated values significantly change the final padding delay).
Figure~\ref{fig:evolved:june} is in a sense similar to
Figure~\ref{fig:evolved:february}; we never get past the initial phase of being
dominated by our padding limits. 

\section{Manually Tuned Machines}
\label{sec:tuned}
Using the evolved machines as a starting point, we set out to create manually
tuned machines. The source code of the evolved machines are available
online\footnote{\url{https://github.com/pylls/padding-machines-for-tor/tree/master/machines/phase2}},
omitted here for sake of brevity.

We used April as a starting point for the \emph{Spring} machine, name from the
fact that April was the most effective evolved machine from spring. The
following changes were made:

\begin{itemize}
	\item upped padding limit to 1500
	\item removed transitions and states that never could do anything
	\item removed useless IAT dists (\texttt{CIRCPAD\_DIST\_NONE})
	\item removed each \texttt{max\_length}, hardly ever hit
\end{itemize}

The main change in behavior of the machine was the upped padding limit, reducing
max recall from 0.57 to 0.47. The full source code of Spring is in
Appendix~\ref{appendix:spring}.

The cleaned up Spring machine in turn was the basis for our final machine:
\emph{Interspace}. The name comes from the song Interspace by Starcadian, that
our Spotify radio played at the time of creation. Thank you randomness (or
rather, deep learning based recommendation engine trained with over ten years of
personal listening history). For Interspace we introduce the concept of
\emph{probabilistically defined} machines. As part of little-t tor startup,
\texttt{circpad\_machines\_init()} initializes all padding machines. Here, we
make the actual definition of Interspace randomized. This is similar to how
ScrambleSuit (and later obfs4) picks a ``polymorphic shape'' per server
instance.

\begin{figure*}[ht]
	\centering
	\includegraphics[width=0.75\textwidth]{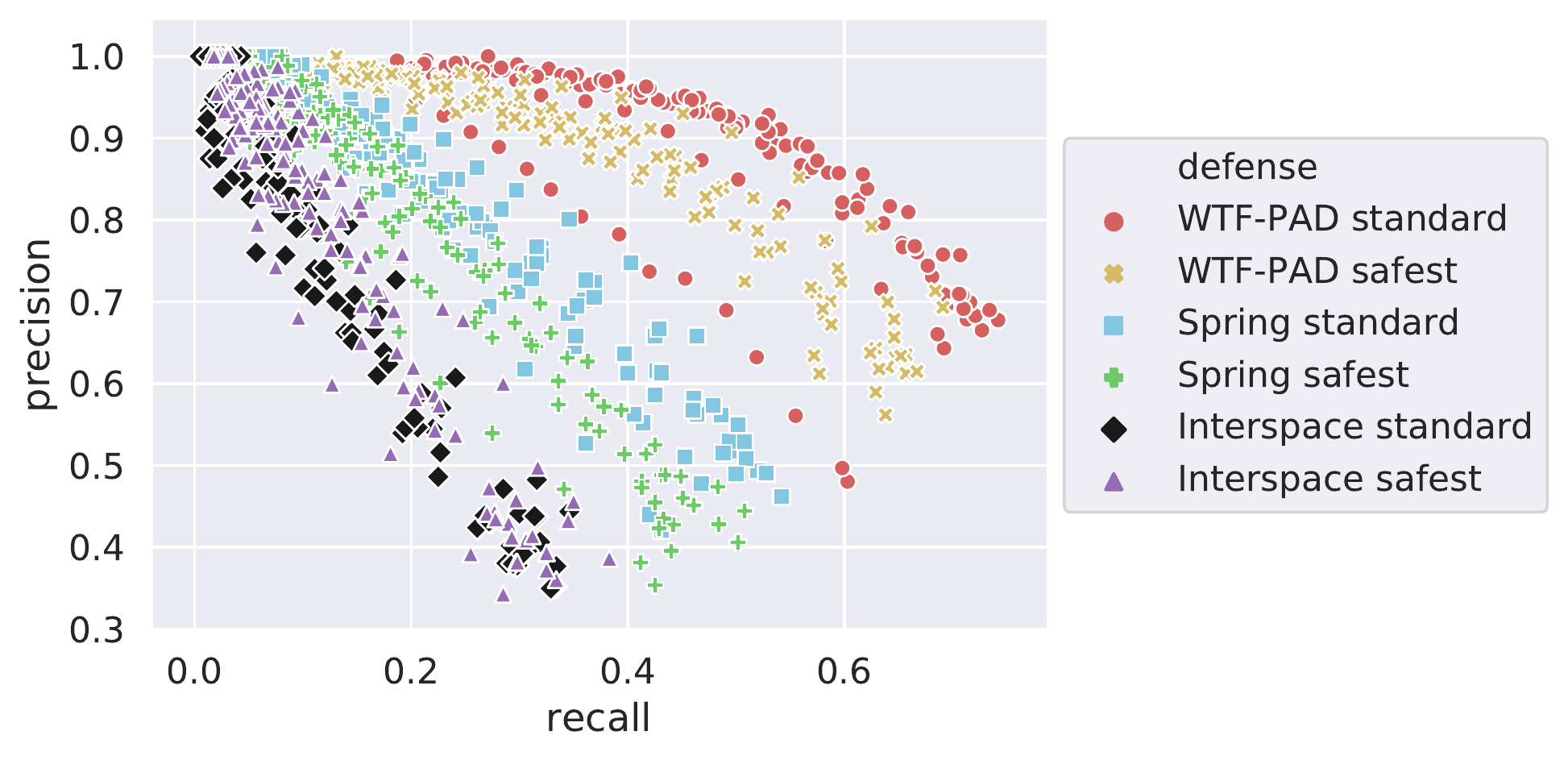}
	\caption{Precision-recall curve of our manually tuned machines, compared to
	the WTF-PAD defense (the norm-rcv config, the best config on our datasets),
	for two security levels from Goodenough February.}
	\label{fig:manual}
\end{figure*}

We spent a modest amount of time (about a week) exploring different
randomization tactics. The full source code of Interspace is in
Appendix~\ref{appendix:interspace}, and has the following main changes from
Spring:

\begin{itemize}
	\item The client machine is changed to include---with 50\% probability
	each---transitions between the two main states that act on received padding
	or non-padding, respectively.
	\item With 50\% probability, the relay machine consists of either the Spring
	relay machine or another, completely hand-crafted, machine. The hand-crafted
	machine, in turn, uniformly selects one of two tactics: either attempt to
	extend real bursts of cells or inject completely fake bursts of modest size.
	For iat and length distributions, we randomize the parameters to the used
	log-logistic (waiting time before injecting a fake burst) and pareto (for
	length, i.e., number of padding cells to send at most, and short iat between
	cells of a burst) distributions.
\end{itemize}
Figure~\ref{fig:manual} shows the effectiveness of our our manually tuned
machines for two security levels. We also include a comparison with one
configuration of WTF-PAD using the most effective default configuration. To make
the WTF-PAD configuration to produce approximately the same amount of padding as
reported by Juarez et al. (177\%), we scaled the timestamps of all cells in the
dataset until we reached nearly the same overhead on the standard dataset
(178\%), as measured by their tool. In gist, WTF-PAD offers some protection,
Spring is notably more effective, and Interspace even more so. The differences
between the defenses are bigger than the differences between Tor Browser
security levels. Table~\ref{table:tuned:overad} shows the overhead for the
defenses and security levels. In a sense, the results are a reverse of
Figure~\ref{fig:manual}: the more effective the defense, the less efficient it
is in comparison.

\begin{table*}[h]
	\caption{Overheads for the two tuned machines as well as WTF-PAD (as
	measured by their own tool) on the February Goodenough dataset for two
	security levels of Tor Browser.}
	\centering
	\label{table:tuned:overad}
	\begin{tabular}{l c c c c}
	\textbf{Machine} & \textbf{Level} & \textbf{Total BW} & \textbf{Sent BW
	(total)} & \textbf{Recv BW (total)} \\
	\midrule 
	Spring & standard & 210\% & 408\% (19\%) & 189\% (81\%) \\
	Spring & safest & 285\% & 731\% (25\%) & 236\% (75\%) \\
	Interspace & standard & 230\% & 604\% (27\%) & 188\% (73\%) \\
	Interspace & safest & 305\% & 909\% (30\%) & 238\% (70\%) \\
	WTF-PAD & standard & 178\% & --- & --- \\
	WTF-PAD & safest & 248\% & --- & --- \\
	\end{tabular}
\end{table*}

\begin{figure*}[h]
	\centering
	\includegraphics[width=0.4\textwidth]{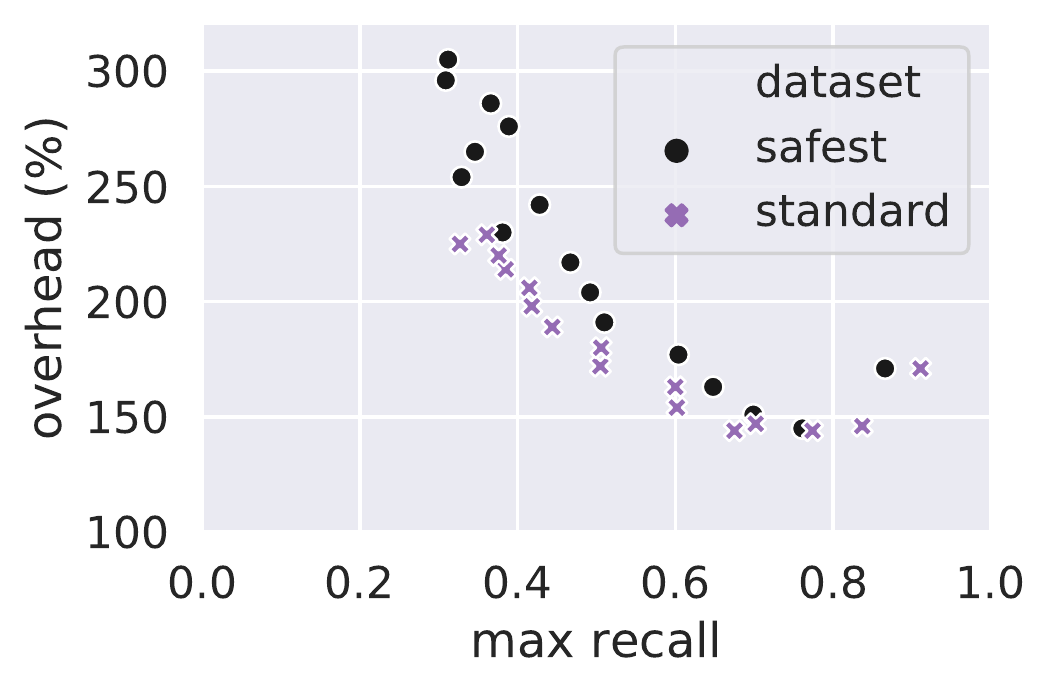}
	\caption{The maximum recall and overhead trade-off of Interspace when
	alternating \texttt{allowed\_padding\_count} from 1 to 1,500 (inclusive) for
	two security levels from Goodenough February.}
	\label{fig:tweakable}
\end{figure*}

The overhead of each machine is capped by \texttt{allowed\_padding\_count} and
\texttt{max\_padding\_percent}, set for both the relay and client machines.
Keeping \texttt{max\_padding\_percent} fixed at 50\%, Figure~\ref{fig:tweakable}
shows the \emph{max-recall} and overhead trade-off of Interspace when
alternating \texttt{allowed\_padding\_count} from 1 to 1,500 (symmetrically, for
both the client and relay) for the standard and safest security levels. For
reference, WTF-PAD had a max call at about 0.7 (average over all folds) at 178\%
overhead for the standard level and a max recall around 0.65 at 248\% overhead
for the safest level. Interspace can reach the same defense effectiveness at
150\% overhead, significantly more efficiently.

\section{Discussion}
\label{sec:discussion}
We consider the good, the bad, and the ugly of our evolved and manually tuned
machines. 

\subsection{The Good}
\label{sec:discussion:good}
In April, we evolved a machine that is a more effective defense than WTF-PAD,
within the constraints of the circuit padding framework of Tor against the
state-of-the-art DF attack. We also showed that WTF-PAD may offer \emph{some}
protection against DF in our more realistic setting---represented by our dataset
and evaluation method---assuming the way we manually tweaked the dataset to make
WTF-PAD's default configuration work was not too drastic. This is an indication
that we need to further consider how we evaluate WF attacks and defenses.

Our manually tuned machines offer further improvements. Interspace provides an
effective defense for both the standard and safest security levels in our
dataset. The overhead is larger than WTF-PAD, so Interspace is less efficient.
That said, the notion of overhead is worth to consider. With
\texttt{allowed\_padding\_count} set to 1500, this is a minimum of 1500 cells:
about 750 KiB. This is not overwhelming. We know from Figure~\ref{fig:zero} that
the early parts of traces are the most important. Accepting, say, a 1 MiB
overhead for web circuits (with added logic on the client to only negotiate the
machine when a stream is attached using port 80 or 443) would likely go a long
way in making attacks harder, yet, not adding significant overheads over typical
website sessions that likely contain media such as images or even videos of
relatively large sizes.

Thanks to the circuit padding framework, Interspace is also tweakable, offering
efficiency--effectiveness trade-offs. We see a clear relationship between
overhead and maximum recall in Figure~\ref{fig:tweakable} for both security
levels. In part, by probabilistically defining the machine, its effectiveness
was improved. This appears to be a promising tactic to further explore.

\subsection{The Bad}
\label{sec:discussion:bad}
In gist, by using genetic programming to evolve machines, our results are
generated by trial-and-error. There is probably a significant room for
improvement. In a sense, we do not really understand why the machines work other
than more general observations such as the relative importance of padding early
rather than later. 

While our dataset was captured on the live network from clients, the relay
portions were simulated. Further, we simulated all our defenses on those traces
and used DF that ignores time. Validating the machines on the live network is a
natural next step. An attenuating circumstance is that each trace in our dataset
was collected on its own unique circuit with no fixed relays. Hopefully, this
means that we have not overfitted our parameters to any particular network
circumstance. Future work will tell.

\subsection{The Ugly}
\label{sec:discussion:ugly}
If the goal was a paper, we are pretty much done now. However, towards the goal
of effective and efficient padding machines for Tor, and not only for
publication, we look deeper. What has hopefully been clear so far is that our
machines \emph{reduce} the effectiveness of the DF attack. The question of what
exactly an ``effective'' defense against WF attacks looks like though has been
debated in the community. Issues such as dataset freshness, different attacker
goals (e.g., inline classification for censorship or ex-post for detection),
website vs website fingerprinting, and victim as well as website base rates are
all good examples. In part, these issues relate to the proper experiment design
as part of the method of evaluation of WF attacks and defenses. We provided some
improvements by taking into account webpage-to-website fingerprinting and
different security levels of Tor Browser.

\begin{figure*}[ht]
	\centering
	\includegraphics[width=0.4\textwidth]{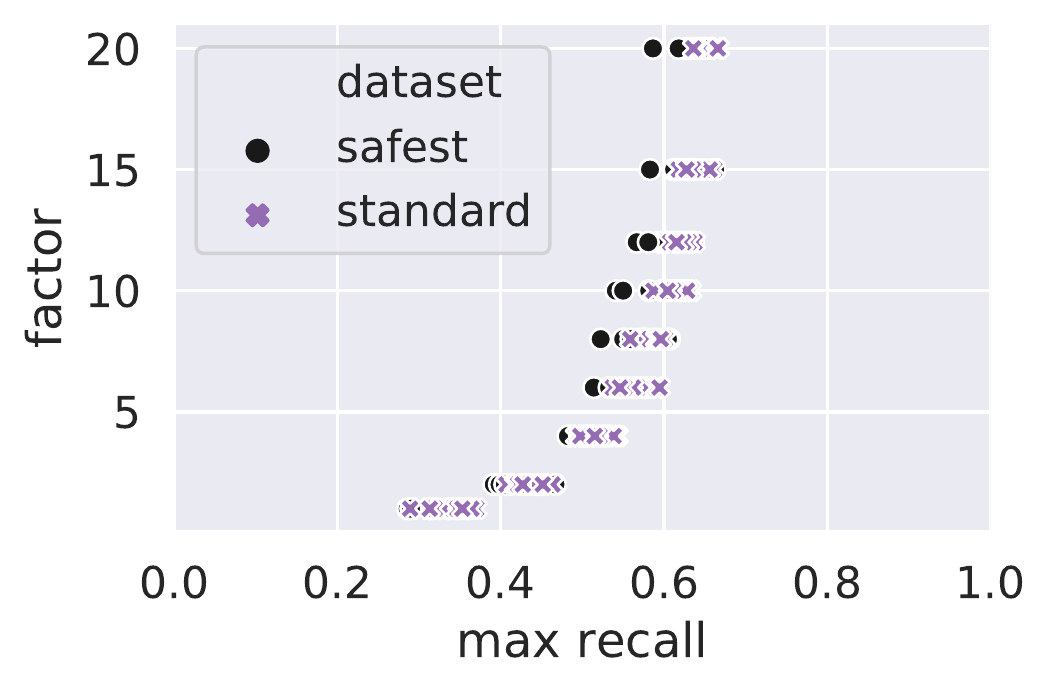}
	\caption{The maximum recall for Interspace as we grow the dataset by a
	variable factor through repeated simulation, for two security levels from
	Goodenough February.}
	\label{fig:multiply}
\end{figure*}

In our experiments we simulate the machines by using circpad-sim. We, as well as
an attacker, can of course simulate each trace multiple times instead of just
once. This should be about the same as visiting webpages of target websites more
times for the sake of training, with the caveat that we also give repeated
samples of the unmonitored websites\footnote{This is not strictly in line with
how unmonitored websites are represented in WF datasets, but because of the
large number of websites labeled as unmonitored (10,000), we believe that this
is negligible.}. By multiplying the number of times we simulate our machines per
trace, we get increasingly more training data. Given that Interspace is
probabilistically defined, simulating the same trace several times may also be
particularly useful. Figure~\ref{fig:multiply} shows the max recall of DF (only
one fold) as we simulate each trace by a factor from 1--20 (from 20,000 to
400,000 samples in the dataset, split 8:1:1). We see that this decreases the
effectiveness of Interspace, improving max recall up to a bit over 0.6 until the
gain stalls around a factor 15 or so.

\section{Conclusions}
\label{sec:conclusions}
The WF setting is challenging. We improved the method of evaluation by
considering webpage-to-website fingerprinting and the security level of Tor
Browser, collecting extensive datasets, and identifying the general parts of
traces most useful for distinguishing between websites. For attacks we used Deep
Fingerprinting (DF), a state-of-the-art deep-learning model.

Using genetic programming and months of wall-time computing with a GPU, we
evolved defenses against DF. The defenses, in the form of padding machines in
Tor's circuit padding framework, were evaluated against DF withing the circuit
padding simulator. Using the most promising evolved machines, we manually tuned
and built new machines, landing in our best candidate: Interspace.

We have created several padding machines for Tor that are more effective
defenses than WTF-PAD, the defense that inspired the design of Tor's circuit
padding framework. Some of our machines are less efficient. Interspace provides
a tweakable trade-off between efficiency and effectiveness, using the global
machine parameters of Tor's framework.

Finally, we showed that the effectiveness of Interspace could be reduced by
significantly increasing the training data available to DF through repeated
simulation of its defense on our collected traces. While it is promising that
this peaked around a factor 15 or so, a max recall of above 0.6 may or may not
be effective enough of a defense. It is also frustrating that central to our
approach is treating DF as a black box. We do not really understand why defenses
work to the extent they do or why, in this particular instance, DF could be
improved to such an extent by adding more training data. Clearly, it is a
challenge to perform genetic programming on such large datasets.

One particularly important lesson is that we observed significantly reduced
attack effectiveness when training on one type of dataset and classifying on
another. We saw this between the standard and safest security levels of TB.
Interspace is based on similar ideas by probabilistically defining its machines.
Table~\ref{table:cross:machines} shows cross-classification across different
trained models, security levels, and datasets. Here we see again that the DF
attack suffers on not being able to train on the correct dataset. This might
appear obvious at first, but it is a largely unexplored area for WF defenses.

\begin{table}[h]
	\caption{Max recall for cross-classification between different trained
	 models and testing datasets. The trained models and tested datasets are
	 from particular defenses and Tor Browser security levels. Red cells
	 highlight where the model and testing dataset match, and orange the
	 second-highest max-recall.}
	\label{table:cross:machines}
	\begin{tabular}{l l l c c c c c c}
	& & & \multicolumn{6}{c}{tested dataset} \\
	& & & \multicolumn{2}{c}{\textbf{Interspace}} &
	\multicolumn{2}{c}{\textbf{Spring}} & \multicolumn{2}{c}{\textbf{WTF-PAD}}
	\\
	& & & standard & safest & standard & safest & standard & safest \\
	\multirow{6}{*}{\STAB{\rotatebox[origin=c]{90}{trained model}}} &
	\multirow{2}{*}{\textbf{Interspace}} & standard & \cellcolor{amaranth!75}
	0.35 & 0.06 & \cellcolor{tangerine!50} 0.30 & 0.05 & 0.03 & 0.01 \\
	& & safest & 0.06 & \cellcolor{amaranth!75} 0.31 & 0.05 & \cellcolor{tangerine!50} 0.26 & 0.01 & 0.02 \\
	& \multirow{2}{*}{\textbf{Spring}} & standard & \cellcolor{tangerine!50}
	0.16 & 0.04 & \cellcolor{amaranth!75} 0.46 & 0.07 & 0.01 & 0.01 \\
	& & safest & 0.04 & \cellcolor{tangerine!50} 0.14 & 0.09 &
	\cellcolor{amaranth!75} 0.42 & 0.02 & 0.02 \\
	& \multirow{2}{*}{\textbf{WTF-PAD}} & standard & 0.05 & 0.03 & 0.07 & 0.03 &
	\cellcolor{amaranth!75} 0.72 & \cellcolor{tangerine!50} 0.14 \\
	& & safest & 0.03 & 0.05 & 0.04 & 0.06 & 0.17 \cellcolor{tangerine!50} &
	\cellcolor{amaranth!75} 0.69 \\
	\end{tabular}
\end{table}

The above begs the question: what is the proper way of evaluating WF defenses?
This has been discussed for a long time in the research community, and while we
hope that Interspace and the lessons learned here move us forward towards the
ultimate goal of effective and efficient padding machines for Tor, we are far
from done.
\\

Code and datasets available at \url{https://github.com/pylls/padding-machines-for-tor}.

\subsection*{Acknowledgements}
This research was funded by generous grants from
\href{https://nlnet.nl/PET/}{NGI Zero PET} and the
\href{https://internetstiftelsen.se/en/}{Swedish Internet Foundation}. Thank you
Mahdi Akil for the review.
\eject
\bibliographystyle{abbrv}
\bibliography{ref}

\begin{thebibliography}{10}

\bibitem{DBLP:journals/popets/BhatLKD19}
S.~Bhat, D.~Lu, A.~Kwon, and S.~Devadas.
\newblock Var-cnn: {A} data-efficient website fingerprinting attack based on
  deep learning.
\newblock {\em Proc. Priv. Enhancing Technol.}, 2019(4):292--310, 2019.

\bibitem{touching}
X.~Cai, X.~C. Zhang, B.~Joshi, and R.~Johnson.
\newblock Touching from a distance: website fingerprinting attacks and
  defenses.
\newblock In {\em {CCS}}, 2012.

\bibitem{cheng1998traffic}
H.~Cheng and R.~Avnur.
\newblock Traffic analysis of {SSL} encrypted web browsing.
\newblock {\em Project paper, University of Berkeley}, 1998.

\bibitem{tor}
R.~Dingledine, N.~Mathewson, and P.~F. Syverson.
\newblock Tor: The second-generation onion router.
\newblock In {\em {USENIX} Security}, 2004.

\bibitem{HerrmannWF09}
D.~Herrmann, R.~Wendolsky, and H.~Federrath.
\newblock Website fingerprinting: attacking popular privacy enhancing
  technologies with the multinomial na{\"{\i}}ve-bayes classifier.
\newblock In {\em {CCSW}}, 2009.

\bibitem{Hintz02}
A.~Hintz.
\newblock Fingerprinting websites using traffic analysis.
\newblock In {\em {PET}}, 2002.

\bibitem{JuarezIPDW16}
M.~Ju{\'{a}}rez, M.~Imani, M.~Perry, C.~D{\'{\i}}az, and M.~Wright.
\newblock Toward an efficient website fingerprinting defense.
\newblock In I.~G. Askoxylakis, S.~Ioannidis, S.~K. Katsikas, and C.~A.
  Meadows, editors, {\em Computer Security - {ESORICS} 2016 - 21st European
  Symposium on Research in Computer Security, Heraklion, Greece, September
  26-30, 2016, Proceedings, Part {I}}, volume 9878 of {\em Lecture Notes in
  Computer Science}, pages 27--46. Springer, 2016.

\bibitem{DBLP:conf/ccs/LiberatoreL06}
M.~Liberatore and B.~N. Levine.
\newblock Inferring the source of encrypted {HTTP} connections.
\newblock In {\em {CCS}}, 2006.

\bibitem{featurediscovery}
N.~Mathews, P.~Sirinam, and M.~Wright.
\newblock Understanding feature discovery in website fingerprinting attacks.
\newblock In {\em IEEE Western New York Image and Signal Processing Workshop
  (WNYISPW)}, 2018.

\bibitem{DBLP:journals/popets/OhSH19}
S.~E. Oh, S.~Sunkam, and N.~Hopper.
\newblock p1-fp: Extraction, classification, and prediction of website
  fingerprints with deep learning.
\newblock {\em Proc. Priv. Enhancing Technol.}, 2019(3):191--209, 2019.

\bibitem{cumul}
A.~Panchenko, F.~Lanze, J.~Pennekamp, T.~Engel, A.~Zinnen, M.~Henze, and
  K.~Wehrle.
\newblock Website fingerprinting at internet scale.
\newblock In {\em {NDSS}}, 2016.

\bibitem{PanchenkoNZE11}
A.~Panchenko, L.~Niessen, A.~Zinnen, and T.~Engel.
\newblock Website fingerprinting in onion routing based anonymization networks.
\newblock In {\em {WPES}}, 2011.

\bibitem{wfwo}
T.~Pulls and R.~Dahlberg.
\newblock Website fingerprinting with website oracles.
\newblock {\em PETS}, 2020.

\bibitem{df}
P.~Sirinam, M.~Imani, M.~Ju{\'{a}}rez, and M.~Wright.
\newblock Deep fingerprinting: Undermining website fingerprinting defenses with
  deep learning.
\newblock In {\em {CCS}}, 2018.

\bibitem{DBLP:conf/sp/SunSWRPQ02}
Q.~Sun, D.~R. Simon, Y.~Wang, W.~Russell, V.~N. Padmanabhan, and L.~Qiu.
\newblock Statistical identification of encrypted web browsing traffic.
\newblock In {\em {IEEE S\&P}}, 2002.

\bibitem{DBLP:conf/wpes/WinterPF13}
P.~Winter, T.~Pulls, and J.~Fu{\ss}.
\newblock Scramblesuit: a polymorphic network protocol to circumvent
  censorship.
\newblock In A.~Sadeghi and S.~Foresti, editors, {\em Proceedings of the 12th
  annual {ACM} Workshop on Privacy in the Electronic Society, {WPES} 2013,
  Berlin, Germany, November 4, 2013}, pages 213--224. {ACM}, 2013.

\end{thebibliography}

\appendix
\section{Spring}
\label{appendix:spring}
\subsection{Client}
\begin{lstlisting}[language=C]
circpad_machine_spec_t *client = tor_malloc_zero(sizeof(circpad_machine_spec_t));
client->conditions.state_mask = CIRCPAD_CIRC_STREAMS;
client->conditions.purpose_mask = CIRCPAD_PURPOSE_ALL;
client->conditions.reduced_padding_ok = 1;

client->name = "spring_client";
client->machine_index = 0;
client->target_hopnum = 2;
client->is_origin_side = 1;
client->allowed_padding_count = 1500;
client->max_padding_percent = 50;
circpad_machine_states_init(client, 3);

client->states[0].next_state[CIRCPAD_EVENT_PADDING_RECV] = 1;

client->states[1].next_state[CIRCPAD_EVENT_NONPADDING_RECV] = 2;

client->states[2].length_dist.type = CIRCPAD_DIST_PARETO;
client->states[2].length_dist.param1 = 4.776842508009852;
client->states[2].length_dist.param2 = 4.807709366988267;
client->states[2].start_length = 1;
client->states[2].iat_dist.type = CIRCPAD_DIST_PARETO;
client->states[2].iat_dist.param1 = 3.3391870088596;
client->states[2].iat_dist.param2 = 7.179045336148708;
client->states[2].dist_max_sample_usec = 9445;
client->states[2].next_state[CIRCPAD_EVENT_NONPADDING_SENT] = 1;
client->states[2].next_state[CIRCPAD_EVENT_PADDING_SENT] = 2;

client->machine_num = smartlist_len(origin_padding_machines);
circpad_register_padding_machine(client, origin_padding_machines);
\end{lstlisting}
\subsection{Relay}
\begin{lstlisting}[language=C]
circpad_machine_spec_t *relay = tor_malloc_zero(sizeof(circpad_machine_spec_t));

relay->name = "spring_relay";
relay->machine_index = 0;
relay->target_hopnum = 2;
relay->is_origin_side = 0;
relay->allowed_padding_count = 1500;
relay->max_padding_percent = 50;
circpad_machine_states_init(relay, 4);
relay->states[0].iat_dist.type = CIRCPAD_DIST_PARETO;
relay->states[0].iat_dist.param1 = 5.460653840184872;
relay->states[0].iat_dist.param2 = 7.080387541173288;
relay->states[0].dist_max_sample_usec = 94722;
relay->states[0].next_state[CIRCPAD_EVENT_NONPADDING_RECV] = 1;
relay->states[0].next_state[CIRCPAD_EVENT_PADDING_RECV] = 1;

relay->states[1].iat_dist.type = CIRCPAD_DIST_LOGISTIC;
relay->states[1].iat_dist.param1 = 1.2767765551835941;
relay->states[1].iat_dist.param2 = 0.11492671368700358;
relay->states[1].dist_max_sample_usec = 31443;
relay->states[1].next_state[CIRCPAD_EVENT_NONPADDING_SENT] = 2;

relay->states[2].length_dist.type = CIRCPAD_DIST_LOGISTIC;
relay->states[2].length_dist.param1 = 4.11964473793041;
relay->states[2].length_dist.param2 = 2.7250362139341764;
relay->states[2].start_length = 5;
relay->states[2].iat_dist.type = CIRCPAD_DIST_LOGISTIC;
relay->states[2].iat_dist.param1 = 5.232180204916029;
relay->states[2].iat_dist.param2 = 5.469677647300559;
relay->states[2].dist_max_sample_usec = 94733;
relay->states[2].next_state[CIRCPAD_EVENT_PADDING_SENT] = 2;
relay->states[2].next_state[CIRCPAD_EVENT_PADDING_RECV] = 3;

relay->states[3].length_dist.type = CIRCPAD_DIST_LOG_LOGISTIC;
relay->states[3].length_dist.param1 = 1.6167675237934875;
relay->states[3].length_dist.param2 = 6.128003159320049;
relay->states[3].start_length = 5;
relay->states[3].iat_dist.type = CIRCPAD_DIST_UNIFORM;
relay->states[3].iat_dist.param1 = 4.270468437086448;
relay->states[3].iat_dist.param2 = 7.926284402139126;
relay->states[3].dist_max_sample_usec = 55878;
relay->states[3].next_state[CIRCPAD_EVENT_NONPADDING_RECV] = 3;
relay->states[3].next_state[CIRCPAD_EVENT_NONPADDING_SENT] = 0;
relay->states[3].next_state[CIRCPAD_EVENT_PADDING_RECV] = 2;

relay->machine_num = smartlist_len(relay_padding_machines);
circpad_register_padding_machine(relay, relay_padding_machines);
\end{lstlisting}

\section{Interspace}
\label{appendix:interspace}
\subsection{Client}
\begin{lstlisting}[language=C]
const struct uniform_t my_uniform = {
    .base = UNIFORM(my_uniform),
    .a = 0.0,
    .b = 1.0,
};

circpad_machine_spec_t *client = tor_malloc_zero(sizeof(circpad_machine_spec_t));
client->conditions.state_mask = CIRCPAD_CIRC_STREAMS;
client->conditions.purpose_mask = CIRCPAD_PURPOSE_ALL;
client->conditions.reduced_padding_ok = 1;

client->name = "interspace_client";
client->machine_index = 0;
client->target_hopnum = 2;
client->is_origin_side = 1;
client->allowed_padding_count = 1500;
client->max_padding_percent = 50;
circpad_machine_states_init(client, 3);

// wait until the relay is active
client->states[0].next_state[CIRCPAD_EVENT_PADDING_RECV] = 1;

// wait for something to either mask the length of or inject a fake request
client->states[1].next_state[CIRCPAD_EVENT_NONPADDING_RECV] = 2;
if (dist_sample(&my_uniform.base) < 0.5) {
   client->states[1].next_state[CIRCPAD_EVENT_PADDING_RECV] = 2;
}

client->states[2].length_dist.type = CIRCPAD_DIST_PARETO;
client->states[2].length_dist.param1 = 4.7;
client->states[2].length_dist.param2 = 4.8;
client->states[2].start_length = 1;
client->states[2].iat_dist.type = CIRCPAD_DIST_PARETO; // tweak for log-logistic?
client->states[2].iat_dist.param1 = 3.3;
client->states[2].iat_dist.param2 = 7.2;
client->states[2].dist_max_sample_usec = 9445;
client->states[2].next_state[CIRCPAD_EVENT_NONPADDING_SENT] = 1;
client->states[2].next_state[CIRCPAD_EVENT_PADDING_SENT] = 2;
if (dist_sample(&my_uniform.base) < 0.5) {
   client->states[2].next_state[CIRCPAD_EVENT_NONPADDING_RECV] = 1;
}

client->machine_num = smartlist_len(origin_padding_machines);
circpad_register_padding_machine(client, origin_padding_machines);
\end{lstlisting}
\subsection{Relay}
\begin{lstlisting}[language=C]
circpad_machine_spec_t *relay = tor_malloc_zero(sizeof(circpad_machine_spec_t));

// short define for sampling uniformly random [0,1.0]
const struct uniform_t my_uniform = {
    .base = UNIFORM(my_uniform),
    .a = 0.0,
    .b = 1.0,
};
#define CIRCPAD_UNI_RAND (dist_sample(&my_uniform.base))

// uniformly random select a distribution parameters between [0,10]
#define CIRCPAD_RAND_DIST_PARAM1 (CIRCPAD_UNI_RAND*10)
#define CIRCPAD_RAND_DIST_PARAM2 (CIRCPAD_UNI_RAND*10)

relay->name = "interspace_relay";
relay->machine_index = 0;
relay->target_hopnum = 2;
relay->is_origin_side = 0;
relay->allowed_padding_count = 1500;
relay->max_padding_percent = 50;
circpad_machine_states_init(relay, 4);

if (CIRCPAD_UNI_RAND < 0.5) {
    /*
    machine has following states:
    0. init: don't waste time early
    1. wait: either extend or fake
    2. extend: obfuscate length of existing bursts
    3. fake: inject fake bursts
    */

    // wait for client to send something, no point in doing stuff too early
    relay->states[0].next_state[CIRCPAD_EVENT_NONPADDING_RECV] = 1;

    if (CIRCPAD_UNI_RAND < 0.5) {
        // wait: extend real burst
        relay->states[1].next_state[CIRCPAD_EVENT_NONPADDING_SENT] = 2;
    } else {
        // wait: inject a fake burst after a while (FIXME: too long below)
        relay->states[1].iat_dist.type = CIRCPAD_DIST_LOG_LOGISTIC;
        relay->states[1].iat_dist.param1 = CIRCPAD_UNI_RAND*1000; // alpha, scale and mean
        relay->states[1].iat_dist.param2 = CIRCPAD_UNI_RAND*10000; // shape, when > 1 larger reduces dispersion
        relay->states[1].dist_max_sample_usec = 100000;
        relay->states[1].next_state[CIRCPAD_EVENT_PADDING_SENT] = 3;
    }

    // extend: add fake padding for real bursts
    relay->states[2].length_dist.type = CIRCPAD_DIST_PARETO;
    relay->states[2].length_dist.param1 = CIRCPAD_RAND_DIST_PARAM1;
    relay->states[2].length_dist.param2 = CIRCPAD_RAND_DIST_PARAM2;
    relay->states[2].start_length = 1;
    relay->states[2].iat_dist.type = CIRCPAD_DIST_PARETO;
    relay->states[2].iat_dist.param1 = CIRCPAD_RAND_DIST_PARAM1;
    relay->states[2].iat_dist.param2 = CIRCPAD_RAND_DIST_PARAM2;
    relay->states[2].dist_max_sample_usec = 10000;
    relay->states[2].next_state[CIRCPAD_EVENT_NONPADDING_SENT] = 1;
    relay->states[2].next_state[CIRCPAD_EVENT_PADDING_SENT] = 2;

    // fake: inject completely fake bursts
    relay->states[3].length_dist.type = CIRCPAD_DIST_PARETO;
    relay->states[3].length_dist.param1 = CIRCPAD_RAND_DIST_PARAM1;
    relay->states[3].length_dist.param2 = CIRCPAD_RAND_DIST_PARAM2;
    relay->states[3].start_length = 4;
    relay->states[3].iat_dist.type = CIRCPAD_DIST_PARETO;
    relay->states[3].iat_dist.param1 = CIRCPAD_RAND_DIST_PARAM1;
    relay->states[3].iat_dist.param2 = CIRCPAD_RAND_DIST_PARAM2;
    relay->states[3].dist_max_sample_usec = 10000;
    relay->states[3].next_state[CIRCPAD_EVENT_NONPADDING_SENT] = 1;
    relay->states[3].next_state[CIRCPAD_EVENT_PADDING_SENT] = 3;
} else {
    // spring-mr
    relay->states[0].iat_dist.type = CIRCPAD_DIST_LOG_LOGISTIC;
    relay->states[0].iat_dist.param1 = CIRCPAD_RAND_DIST_PARAM1;
    relay->states[0].iat_dist.param2 = CIRCPAD_RAND_DIST_PARAM2;
    relay->states[0].dist_max_sample_usec = 10000;
    relay->states[0].next_state[CIRCPAD_EVENT_NONPADDING_RECV] = 1;
    relay->states[0].next_state[CIRCPAD_EVENT_PADDING_RECV] = 1;

    relay->states[1].iat_dist.type = CIRCPAD_DIST_LOG_LOGISTIC;
    relay->states[1].iat_dist.param1 = CIRCPAD_RAND_DIST_PARAM1;
    relay->states[1].iat_dist.param2 = CIRCPAD_RAND_DIST_PARAM2;
    relay->states[1].dist_max_sample_usec = 31443;
    relay->states[1].next_state[CIRCPAD_EVENT_NONPADDING_SENT] = 2;

    relay->states[2].length_dist.type = CIRCPAD_DIST_LOG_LOGISTIC;
    relay->states[2].length_dist.param1 = CIRCPAD_RAND_DIST_PARAM1;
    relay->states[2].length_dist.param2 = CIRCPAD_RAND_DIST_PARAM2;
    relay->states[2].start_length = 5;
    relay->states[2].iat_dist.type = CIRCPAD_DIST_LOG_LOGISTIC;
    relay->states[2].iat_dist.param1 = CIRCPAD_RAND_DIST_PARAM1;
    relay->states[2].iat_dist.param2 = CIRCPAD_RAND_DIST_PARAM2;
    relay->states[2].dist_max_sample_usec = 100000;
    relay->states[2].next_state[CIRCPAD_EVENT_PADDING_SENT] = 2;
    relay->states[2].next_state[CIRCPAD_EVENT_PADDING_RECV] = 3;

    relay->states[3].length_dist.type = CIRCPAD_DIST_LOG_LOGISTIC;
    relay->states[3].length_dist.param1 = CIRCPAD_RAND_DIST_PARAM1;
    relay->states[3].length_dist.param2 = CIRCPAD_RAND_DIST_PARAM2;
    relay->states[3].start_length = 5;
    relay->states[3].iat_dist.type = CIRCPAD_DIST_LOG_LOGISTIC;
    relay->states[3].iat_dist.param1 = CIRCPAD_RAND_DIST_PARAM1;
    relay->states[3].iat_dist.param2 = CIRCPAD_RAND_DIST_PARAM2;
    relay->states[3].dist_max_sample_usec = 55878;
    relay->states[3].next_state[CIRCPAD_EVENT_NONPADDING_RECV] = 3;
    relay->states[3].next_state[CIRCPAD_EVENT_NONPADDING_SENT] = 0;
    relay->states[3].next_state[CIRCPAD_EVENT_PADDING_RECV] = 2;
}

relay->machine_num = smartlist_len(relay_padding_machines);
circpad_register_padding_machine(relay, relay_padding_machines);
\end{lstlisting}

\end{document}